\documentclass{article}

\usepackage[utf8]{inputenc}
\usepackage{amsmath}
\usepackage{amssymb}
\usepackage{latexsym}
\usepackage{graphicx}
\usepackage{pgfplots}
  \pgfplotsset{width=10cm,compat=1.9}
\usepackage{tikz}
  \tikzset{
    conv/.style=  
     {black,draw=black,fill=conv,
      rectangle,minimum height=1cm}
  }
  \tikzstyle{line} = [draw, -latex']
  \usetikzlibrary{positioning}
  \usetikzlibrary{shapes}
  \usetikzlibrary{arrows}
  \usetikzlibrary{chains}
  \usetikzlibrary{patterns}
  \usetikzlibrary{snakes}
  \usetikzlibrary{calc}
  \usetikzlibrary{spy}
\usepackage{subcaption} 
\usepackage{lscape} 
\usepackage{booktabs}
\usepackage{multirow}
\usepackage{array}
\usepackage[toc,page]{appendix}
\usepackage[linesnumbered,ruled,vlined]
  {algorithm2e}
  \SetKwComment{Comment}{/* }{ */}
\usepackage{acro}
\DeclareAcronym{CT}{
  short = CT,
  long = Computed Tomography}
\DeclareAcronym{SDF}{
  short = SDF,
  long = Self-supervised Denoiser Framework}
\DeclareAcronym{CBCT}{
  short = CBCT,
  long = cone-beam CT}
\DeclareAcronym{AGD}{
  short = AGD,
  long = accelerated gradient descent}
\DeclareAcronym{FDK}{
  short = FDK,
  long = Feldkamp-Davis-Kress}
\DeclareAcronym{FBP}{
  short = FBP,
  long = filtered backprojetion}  
\DeclareAcronym{CNN}{
  short = CNN,
  long = convolutional neural network}
\DeclareAcronym{NN}{
  short = NN,
  long = neural network}
\DeclareAcronym{PSNR}{
  short = PSNR,
  long = peak signal-to-noise ratio}
\DeclareAcronym{GD}{
  short = GD,
  long = gradient descent}
\DeclareAcronym{LS}{
  short = LS,
  long = least-squares}

\usepackage{hyperref}
\hypersetup{%
    unicode=true,     
    colorlinks=true, 
    linkcolor=blue,   
    citecolor=red, 
    filecolor=blue,   
    urlcolor=blue
}  
\usepackage{xcolor}
  \definecolor{conv}{HTML}{1E90FF}
\usepackage{xspace}
\usepackage{todonotes} 

\newcommand{\Real}{\mathbb{R}}
\newcommand{\rand}[1]{\mathsf{#1}}
\newcommand{\FwdOp}{\operatorname{\mathcal{A}}}
\newcommand{\FwdOpInv}{\FwdOp^{\dagger}}
\newcommand{\RecoOp}{\operatorname{\mathcal{R}}}
\newcommand{\MAH}{\operatorname{MAH}}
\newcommand{\MLP}{\operatorname{MLP}}
\newcommand{\nii}{Noise2Inverse\xspace}
\newcommand{\argmin}{\operatorname*{arg\,min}}
\newcommand{\Expect}{\operatorname{\mathbb{E}}}
\newcommand{\Loss}{\operatorname{\mathcal{L}}}
\newcommand{\loss}{\operatorname{\ell}}
\newcommand{\Encoder}{\operatorname{\mathcal{E}}}
\newcommand{\Decoder}{\operatorname{\mathcal{D}}}

\usepackage[backend=biber,maxnames=3, sorting=none]{biblatex}
\title{SDF: A Framework for Self-supervision}
\author{Emilien Valat \and Andreas Hauptmann \and Ozan Öktem}

\begin{document}
\maketitle

\begin{abstract}
Reconstructing images using \ac{CT} in an industrial context leads to specific challenges that differ from those encountered in other areas, such as clinical \ac{CT}. Indeed, non-destructive testing with industrial \ac{CT} will often involve scanning multiple similar objects while maintaining high throughput, requiring short scanning times, which is not a relevant concern in clinical \ac{CT}. Under-sampling the tomographic data (sinograms) is a natural way to reduce the scanning time at the cost of image quality since the latter depends on the number of measurements. In such a scenario, post-processing techniques are required to compensate for the image artifacts induced by the sinogram sparsity. 

We introduce the \ac{SDF}, a self-supervised training method that leverages pre-training on highly sampled sinogram data to enhance the quality of images reconstructed from under-sampled sinogram data. The main contribution of \ac{SDF} is that it proposes to train an image denoiser in the sinogram space by setting the learning task as the prediction of one sinogram subset from another. As such, it does not require ground-truth image data, leverages the abundant data modality in \ac{CT}, the sinogram, and can drastically enhance the quality of images reconstructed from a fraction of the measurements.

We demonstrate that \ac{SDF} produces better image quality, in terms of \ac{PSNR}, than other analytical and self-supervised frameworks in both 2D fan-beam or 3D \ac{CBCT} settings. Moreover, we show that the enhancement provided by \ac{SDF} carries over when fine-tuning the image denoiser on a few examples, making it a suitable pre-training technique in a context where there is little high-quality image data. Our results are established on experimental datasets, making \ac{SDF} a strong candidate for being the building block of foundational image-enhancement models in \ac{CT}.
\end{abstract}

\acresetall
\section{Introduction}
X-ray \ac{CT} is an imaging technique that computes volumetric images from a set of indirect observations, the sinogram, obtained by sampling an object with an X-ray beam under different viewing angles. \ac{CT} is used routinely in various situations, such as medical imaging and non-destructive testing for industrial applications. The latter setting involves scanning objects while maintaining high throughput, highlighting the need for fast sampling and reconstruction routines. A natural way to limit sampling time is to reduce the number of measurements of each object at the expense of the reconstructed image's quality. Advanced computational techniques are required to accommodate sparse measurements during the reconstruction process or enhance the images in a post-processing step. 

Tomographic image reconstruction is an inverse problem. That is, given observations $y \in Y$, recovering the causal factor $x \in X$ that produced them. For X-ray \ac{CT}, the set of observations is called the \emph{sinogram} and is obtained by sampling an object from different viewing angles. The \emph{data model} is a formalised computational model for generating the sinogram from an image. It can be defined as:
\begin{equation}\label{eq:y_ax}    
    y = \FwdOp x + \epsilon
\end{equation}
where $\FwdOp \colon X \to Y$ is the \emph{forward operator} that models the interaction between the X-rays and the object without observation noise. The term $\epsilon$ represents the latter and is inherent to the sampling process. Although the radiative transport equation accurately models the interaction, it is common in \ac{CT} to consider a simplified model that ignores beam hardening (monochromatic X-rays) and scattering. Under these simplifications and assuming data has undergone suitable pre-processing, one can set the forward operator $\FwdOp$ to the ray transform.

In tomography, reconstructing an image is equivalent to solving \eqref{eq:y_ax}. As such, when $\FwdOp$ is the ray transform, this amounts to its inversion, which is an ill-posed process due to the noise introduced during sampling. Regularisation is then needed to mitigate the solution's intrinsic instability and the noise amplification that comes from the ill-posedness of the inversion. These regularisation methods can be analytical, iterative, variational, or data-driven \cite{arridge_solving_2019}. 
The data-driven approach methods outperform others when sinogram data is missing due to undersampling, is highly noisy, or both, as shown experimentally in \cite{genzel_near-exact_2022, leuschner_quantitative_2021}. Among these, the best-performing methods are often based on domain-adapted neural networks, meaning that they rely on reconstruction operators $\RecoOp_{\theta} \colon Y \to X$ given by deep \acp{NN} with an architecture that combines a handcrafted physics-based model about the forward operator and a learnable block. One other advantage of the data-driven reconstruction networks is that they can be interfaced with preprocessing and post-processing operators for tasks such as image classification \cite{adler_task_2022}, segmentation \cite{valat_empirical_2024} and registration \cite{eijnatten_3d_2021}. However, this integration with other networks can be problematic when the reconstruction operator is trained in a supervised manner. In the most classical case, where a neural network is trained to map a noisy or undersampled sinogram to a high-quality image, the training requires abundant target data, leading to the train networks tending to enforce features in the inferred image like the ones present in the target images they learned from. This, in turn, hinders the interfacing with post-processing operators, for these features might not be adapted inputs to the post-processing operators. As such, there is an interest in considering other learning protocols that do not involve ground-truth images during training. Although such approaches exist \cite{hendriksen_noise2inverse_2020, kosomaa_simulator-based_2023}, they are not designed for reconstructing images for under-sampled sinogram data. 
This paper contributes to this active field of research by providing a self-supervised framework that only relies on sinogram data for training denoisers to enhance images reconstructed from undersampled data.

\section{Related Work}
\Ac{SDF} is a learning technique that leverages the abundant data modality in X-ray \ac{CT}, the sinogram, for pretraining an image denoiser. As such, it relates closely to the self-supervised framework Noise2Inverse \cite{hendriksen_noise2inverse_2020}. 

The underlying idea is to create input-target image pairs by reconstructing them from different sinogram subsets and train a denoiser \ac{NN} in a self-supervised way. \ac{NN}'s training goal is to infer an image reconstructed from one subset from an image reconstructed from a different one. Despite its good performance in removing measurement noise, the Noise2Inverse framework ``does not remove artifacts resulting from under-sampling'' according to its authors. We surmise that the non-local nature of \ac{CT} images prevents an image denoiser trained on images exhibiting under-sampling artifacts from removing them. This supposition is illustrated by the difference in performance between training strategy $1:X$ and $X:1$, highlighting the importance of artifact-free input images. Building on the Noise2Inverse framework, the Proj2Proj framework \cite{unal_proj2proj_2024} investigates the use of projection data corruption and masking to train an image denoiser in a self-supervised fashion in the low-dose \ac{CT} case, but is not designed to enhance images reconstructed from sparse measurements.

Self-supervision is also used successfully by \cite{kosomaa_simulator-based_2023}. They use a loss in the sinogram space to train an image denoiser, producing high-fidelity images for CBCT and helical trajectory. Their approach relies on reconstructing a high-quality image from all but 12 projections and resampling the image at the locations of the 12 ones left aside to minimise a loss in the sinogram space. As such, they compare ``simulated'' projections to the ones sampled initially. Although this approach is sound for medical \ac{CT}, as undersampling is not used in this field, they do not evaluate their method on undersampled data.
Moreover, they use a substantial image denoiser: for 3D \ac{CBCT}, their network totals 1.73M parameters (988k for the sinogram network and 742k for the volume network), while we successfully use a 10k parameters network, removing the uncertainty about the need for large neural networks in our approach. We have also found their results hard to implement, as their method involves custom CUDA kernels, while our method is designed for pure PyTorch \cite{paszke_pytorch_2019} implementation. Loss in the sinogram space is also used in \cite{chen_robust_2022} to enforce so-called robust equivariance. 

Self-supervision is increasingly popular in the \ac{CT} community, as the emerging methods to train image reconstruction NNs without high-quality image targets show. Self-supervision is tightly connected to foundational models, which are general-purpose models pre-trained using self-supervision on data at scale and fine-tuned using supervision on limited data. Such models, like the GPT series from OpenAI or BERT \cite{devlin_bert_2019}, appear in the field of analysis of CT images\cite{hamamci_developing_2024} but are yet to be developed for CT image reconstruction. As such, there is a strong interest in the mathematical formalisation of self-supervision in image reconstruction and the experimental validation of the procedures on real datasets.

Here, we propose a new self-supervised method to train image denoisers for linear inverse problems and develop its mathematical formalisation. We then investigate \ac{SDF}'s performance for two noise settings and compare it with other analytical, iterative and data-driven methods. We further show that \ac{SDF} is a suitable pretraining strategy for few-shot supervised training of high-quality image denoisers for undersampled sinogram data. We also assert the method's robustness against angular sparsity of the sinogram data and its ability to scale to 3D CBCT.

\section{Methods}
We define the reconstruction operator $\RecoOp_{\theta} \colon Y_0 \to X$ for sinograms in some fixed sub-space $Y_0 \subset Y$ as
\begin{equation}\label{eq:SDFRecoOp}
  \RecoOp_{\theta} := \Lambda_{\theta} \circ \FwdOpInv_0.
\end{equation}
In the above, $\Lambda_{\theta} \colon X \to X$ is an image restoration/denoising network, 
$\FwdOpInv_0 \colon Y_0 \to X$ is some handcrafted pseudo inverse of $\FwdOp_0 := \FwdOp \circ \pi_0$ with $\pi_0 \colon Y \to Y_0$ denoting an appropriate (preferably linear) restriction/re-binning operator that maps sinograms in $Y$ to those in $Y_0$.

\subsection{Learning protocol}
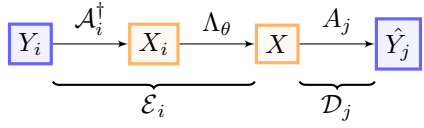
\begin{figure}[h]
    \centering
    \begin{tikzpicture}[
    sinogramnode/.style={rectangle, draw=blue!60, fill=blue!5, very thick, minimum size=5mm},
    imagenode/.style={rectangle, draw=orange!60, fill=orange!5, very thick, minimum size=5mm}
    ]
    \node[sinogramnode] (Yi) {$Y_i$};
    \node[imagenode] (X_i) [right=of Yi] {$X_i$};
    \node[imagenode] (X) [right=of X_i] {$X$};
    \node[sinogramnode] (Yjhat) [right=of X] {$\hat{Y_j}$};
    
    \path[line] (Yi) -- node [above ] {$\FwdOpInv_i$} (X_i);
    \path[line] (X_i) -- node [above ] {$\Lambda_{\theta}$} (X);
    \path[line] (X) -- node [above ] {$A_j$} (Yjhat);

    
    \draw [
    thick,
    decoration={
        brace,
        mirror,
        raise=0.5cm
    },
    decorate
    ] (Yi) -- (X) node [pos=0.5,anchor=north,yshift=-0.55cm] {$\Encoder_i$}; 
    \draw [
    thick,
    decoration={
        brace,
        mirror,
        raise=0.5cm
    },
    decorate
    ] (X) -- (Yjhat) node [pos=0.5,anchor=north,yshift=-0.55cm] {$\Decoder_j$}; 
    \
    
    \end{tikzpicture}
    \caption{The blue nodes represent sinogram-space data, and the orange nodes represent image-space data. The composition of $\FwdOpInv_i$ with $\Lambda_{\theta}$ (denoted by $\Encoder_i$)  and $A_j$ (denoted $\Decoder_j$) can be seen as encoding and decoding parts of a sinogram autoencoder that has $X$ (set of images) as its latent space.}
    \label{fig:full_architecture}
\end{figure}

Learning the reconstruction operator $\RecoOp_{\theta} \colon Y_0 \to X$ in \eqref{eq:SDFRecoOp} is a by-product of learning the image restoration/denoising operator $\Lambda_{\theta} \colon X \to X$, which in turn is typically learned by training against supervised data in $X$. The key idea in \ac{SDF} is to learn the reconstruction operator from self-supervised data in $Y$. We first subdivide the sinogram space $Y$ into disjoint sets, so $Y = Y_{1} \cup \ldots \cup Y_M$. The restrictions of the forward operator and its pseudo-inverse to these sets are then given as
\[
  \FwdOp_i \colon X \to Y_i
  \quad\text{and}\quad
  \FwdOpInv_i \colon Y_i \to X
  \quad\text{for $i=1, \ldots, M$.}
\]
With the above, we can for any pair $i,j = 1, \ldots, M$ define the following sinogram-to-sinogram mapping, as illustrated in Fig.~\ref{fig:full_architecture}:  
\[
  \Gamma_{i,j}^{\theta} \colon Y_i \to Y_j
  \quad\text{where}\quad
  \Gamma_{i,j}^{\theta} := \FwdOp_j \circ \Lambda_{\theta} \circ \FwdOpInv_i.
\]

Once the sinogram-to-sinogram mappings have been defined, one can use $\Gamma_{i,j}^{\theta} \colon Y_i \to Y_j$ to predict a sinogram $y_j \in Y_j$ from a sinogram $y_i \in Y_i$.  To formalise this, we assume sinograms are generated by a $Y$-valued random variable $\rand{y} \sim \mu$. Next, let $\zeta_i \colon Y \to Y_i$ be a re-binning operator that maps a sinogram in $Y$ to one in $Y_i$. Then we can define the $Y_i$-valued random variable $\rand{y}_i := \zeta_i(\rand{y})$ and set-up the following learning problem: 
\begin{equation}\label{eq:LearnProb}
\hat{\theta} \in 
\argmin_{\theta \in \Theta} 
\sum_{\substack{i,j=1 \\ i\neq j}}^M \Expect_{\rand{y}_i,\rand{y}_j}\Bigl[ 
  \loss_{j}\bigl( \Gamma_{i,j}^{\theta}(\rand{y}_i), \rand{y}_j \bigr)
\Bigr]
\quad\text{for some loss $\loss_{j} \colon Y_j \times Y_j \to \Real$.}
\end{equation}

To formulate the empirical counterpart of \eqref{eq:LearnProb}, assume we have unsupervised training data in the form of sinograms $y_1, \ldots, y_m \in Y$ that are i.i.d.\@ samples of $\rand{y}$.
We then use re-binning to generate sinograms in $Y_1, \ldots, Y_M$ by
\[ y^i_l := \zeta_i(y_l) \in Y_i 
   \quad\text{for $i=1,\ldots, M$ and $l=1,\ldots, m$.} 
\]
In particular, for each $i,j=1,\ldots, M$, we can generate pairs of sinograms $(y^i_l,y^j_l) \in Y_i \times Y_j$ for $l=1,\ldots, m$ that are random samples of $(Y_i \times Y_j)$-valued random variable $(\rand{y}_i, \rand{y}_j)$.
One may view these pairs as some form of ``supervised'' sinogram data that can be used for learning $\hat{\theta} \in \Theta$ by minimising 
\[ \theta \mapsto \sum_{\substack{i,j=1 \\ i\neq j}}^M   \Loss_{i,j}(\theta) \]
where the objective $\Loss_{i,j} \colon \Theta \to \Real$ is the empirical counterpart of the expectation in \eqref{eq:LearnProb}: 
\[
  \Loss_{i,j}(\theta) := \frac{1}{m}\sum_{l=1}^m 
  \loss_{j}\bigl( \Gamma_{i,j}^{\theta}(y^i_l), y^{j}_l \bigr)
  \quad\text{for $\theta \in \Theta$ and $i,j=1,\ldots, M$}
\]

For $i=1,\ldots, M-1$ we can in particular learn $\hat{\theta} \in \Theta$ by minimising 
\begin{equation}\label{eq:TotalLossSeq}
    \theta \mapsto \sum_{i=1}^M \Loss_{i,i+1}(\theta).
\end{equation}
One can now use a \ac{GD} scheme to minimise \eqref{eq:TotalLossSeq}, which with step-size $\omega>0$ reads as follows:
\begin{equation}\label{eq:LearnIterScheme}
    \theta_{k+1} = \theta_{k} - \omega \sum_{i=1}^M \nabla \Loss_{i,i+1}(\theta_{k})
    \quad\text{for $k=0,1,\ldots$.}    
\end{equation}
An alternate way that avoids the summation in \eqref{eq:LearnIterScheme} is the following: 
\begin{equation}\label{eq:ValatScheme}
\theta_{k+1} =\theta_{k} - \omega \nabla \Loss_{i_k,i_k+1}(\theta_{k}) 
\quad\text{where $i_k :=k \,(\operatorname{mod}\, M-1)+1$.}
\end{equation}
The advantage of \ref{eq:ValatScheme} is to reduce the memory footprint of the training procedure and will be used throughout the experiments.

\subsection{Comments and remarks}
One can think of \ac{SDF} as training an auto-encoder of which latent space is the image. Autoencoders are neural networks that learn a latent representation of unlabeled data. They comprise an encoder $\Encoder$ that maps the input to a low-dimensional, latent representation $z$ and a decoder $\Decoder$ that transforms the latter back to the input data. When trained to reconstruct a subset of their data given another, these autoencoders are said to be trained in a self-supervised fashion. The design of these networks, i.e. the properties of $\Encoder$, $\Decoder$ and $z$, is more a matter of hyper-parameter tuning than a proper model-informed procedure. As such, when building a sinogram autoencoder to recover image data, we propose that the latent space be an array with dimensions equal to the ones of the volume on which the reconstruction will be computed. Following that hypothesis,  we suggest that $\Encoder$ and $\Decoder$ should entail the $\FwdOpInv$ and $\FwdOp$ operators corresponding to the geometry at hand. We finally put forward the idea of training the neural network in a self-supervised fashion by dividing the sinogram into subsets, allowing the processing of sparse sinogram data. 

As with the sinogram sub-sampling strategy, one could devise several strategies to iterate through the subsets. In this report, we train $\Lambda_{\theta}$ to infer the next closest subset to the one we provide, but inferring less correlated subsets might be relevant given the task at hand. We experiment with orbital subsets for \ac{CBCT} and angular subsets for both \ac{CBCT} and 2D \ac{CT}.

\section{Experiments}

\subsection{Datasets}
We experiment only on datasets containing real measurements. For 2D \ac{CT}, we use the 2DeteCT dataset \cite{kiss_2detect_2023}; for 3D \ac{CT}, we use the Walnuts dataset \cite{der_sarkissian_cone-beam_2019}. The LION \cite{biguri_learned_nodate} toolbox provides pre-processing and data-loading functions for the two datasets. We refer the reader to the original papers for a description of the geometries used in these datasets, which we used as is. 

\subsubsection{2DeteCT dataset}
\begin{figure}
\begin{subfigure}{.5\textwidth}
  \centering
  \includegraphics[width=.8\linewidth]{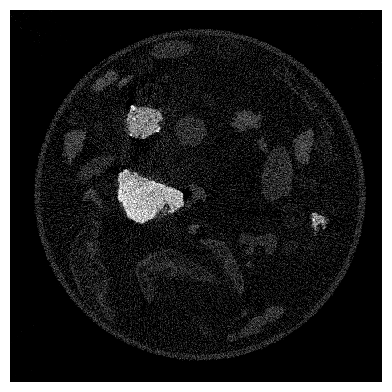}
  \caption{Mode1 Slice}
  \label{fig:mode1_slice}
\end{subfigure}%
\begin{subfigure}{.5\textwidth}
  \centering
  \includegraphics[width=.8\linewidth]{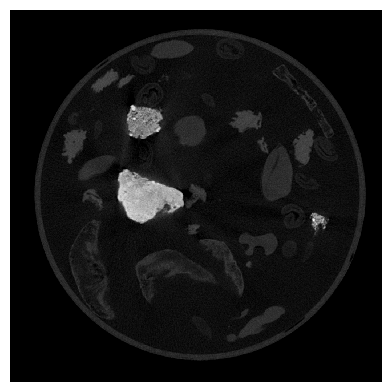}
  \caption{Mode2 Slice}
  \label{fig:mode2_slice}
\end{subfigure}
\caption{Slices of the same object reconstructed from mode1 (\subref{fig:mode1_slice}) and mode2 (\subref{fig:mode2_slice}) sinograms.}
\label{fig:2detect_samples}
\end{figure}

The 2DeteCT dataset contains 5000 2D slices reconstructed by Nesterov Accelerated Gradient Descent associated with sinograms acquired with different modes and fan-beam geometry. The ``\emph{Mode1}'' corresponds to sinograms with low photon count, and the ``\emph{Mode2}'' corresponds to sinograms with normal photon count. Mode1 sinograms can be considered as ``noisy'' and Mode2 as ``clean'', as per the wording of the original paper. Clean sinograms were acquired using a tube power of 90W, whereas their noisy counterparts were acquired using a tube power of 3W. This dataset allows us to experiment on sparse-view \ac{CT} and low-photon count \ac{CT}. Fig.~\ref{fig:2detect_samples} shows a slice of the 2DeteCT dataset acquired with Mode1 and Mode2. 

\subsubsection{Walnuts dataset}
\begin{figure}
\begin{subfigure}{.5\textwidth}
  \centering
  \includegraphics[width=.8\linewidth]{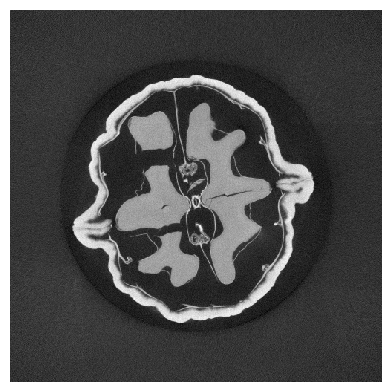}
  \caption{\ac{FDK} + orbit 1}
  \label{fig:fdk_orbit1}
\end{subfigure}%
\begin{subfigure}{.5\textwidth}
  \centering
  \includegraphics[width=.8\linewidth]{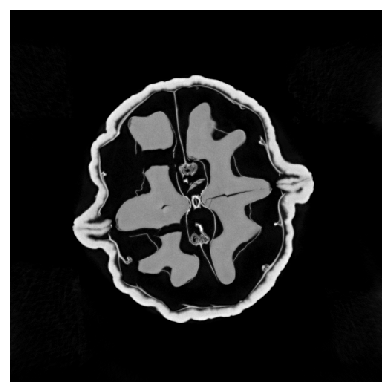}
  \caption{\ac{LS} + all orbits}
  \label{fig:ag_all_orbits}
\end{subfigure}
\caption{Central slice of the same volume reconstructed from \ac{FDK} ran on orbit~1 (\subref{fig:fdk_orbit1}), and AG ran on all orbits (\subref{fig:ag_all_orbits}).}
\label{fig:walnut_sample}
\end{figure}

The Walnut dataset contains 42 tomographic sinograms acquired using a \ac{CBCT} geometry comprising three orbits, noted 1, 2 and 3, with 1200 intensity measurements for each orbit. The corresponding 3D volumes are 501-voxel sized, and they are reconstructed from the over-sampled sinogram data by \ac{LS} reconstruction\footnote{This is minimizing least-squares error in sinogram space without explicit regulariser.} using \ac{AGD}. this dataset allows us to experiment with angular and orbital sparsity. Fig.~\ref{fig:walnut_sample} shows a slice of a volume reconstructed with the \ac{FDK} method from one orbit and least-squares reconstruction from three orbits.

\subsection{Implementation Details}
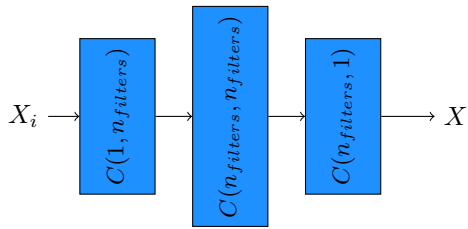
\begin{figure}
  \centering
  \begin{tikzpicture}
    \node (X_i) at (1.25,0) {$X_i$};
  
    \node[conv,rotate=90,minimum width=2cm] (conv1) at (2.5, 0) {\small$C(1,n_{filters})$};
    \node[conv,rotate=90,minimum width=2cm] (conv2) at (4,0) {\small$C(n_{filters},n_{filters})$};
    \node[conv,rotate=90,minimum width=2cm] (conv3) at (5.5,0)    {\small$C(n_{filters},1)$};
  
    \node (X) at (7,0) {$X$};
  
    \draw[->] (X_i) -- (conv1);
    \draw[->] (conv1) -- (conv2);
    \draw[->] (conv2) -- (conv3);
    \draw[->] (conv3) -- (X);
    
  \end{tikzpicture}
  \vskip 6px
  \caption{Architecture of the image denoiser $\Lambda_{\theta}$ used in all our experiments. $C(n,m)$ denotes a convolutional layer with $n$ input layers and $m$ output layers, a kernel size of 5 and a padding of 2, with LeakyReLU activation. For 3D CT, $n_{filters} = 8$ and for 2D CT, $n_{filters} = 32$.}
  \label{fig:vanilla_CNN}
\end{figure}

For a fair comparison between methods, we experiment with a simple $\Lambda_{\theta}$ network, defined as a vanilla-\ac{CNN}, and illustrated in Fig.~\ref{fig:vanilla_CNN}. We use Normal weight initialisation, the Adam optimiser, with a learning rate of $10^{-5}$ and set the batch size to 8 for 2D and 4 for 3D. The training takes 40 epochs over the dataset, split unless mentioned otherwise, between 80\% training, 10\% validation and 10\% testing. We use gradient clipping, as we found it to stabilise training. We use Tomosipo \cite{hendriksen_tomosipo_2021} and its ASTRA \cite{aarle_fast_2016} backend to implement differentiable forward and backward operators.

\subsection{Low Photon Count and Sparse-view \ac{CT}}
This experiment assesses how \ac{SDF} performs at training a denoiser to improve sparse-view images reconstructed from Mode1 and Mode2 sinograms of the 2DeteCT dataset. To do so, we use an angular subsampling of factor 10, reducing the 3600 measurement sinograms to 360 uniformly distributed measurement sinograms. This yields ten angular subsets and $(\FwdOp, \FwdOpInv)$ pairs.

We compare \ac{SDF} to analytic, iterative and self-supervised image reconstruction frameworks, respectively, the \ac{FBP}, the \ac{LS} and the \nii. Our use case is sparse-view \ac{CT}, so we train the \nii using the 1:N method, as described in \cite{hendriksen_noise2inverse_2020}. Our golden standard for image quality is the Mode2, high-quality \ac{LS} reconstruction computed using Nesterov-\acs{AGD}, as described in \cite{kiss_2detect_2023}. Our results are presented in Table~\ref{tab:results_2deteCT_sparse_view}.

\begin{table}[]
    \centering
    \begin{tabular}{cll}
    \toprule
    Mode & Method & Sparse-View \acs{PSNR} \\
    \midrule
    \multirow{4}{*}{1} & \ac{FBP} & $6.77 \pm 2$ \\
    & \ac{LS}               & $26.64 \pm 1.94$ \\
    & \ac{SDF}           & $\mathbf{26.96 \pm 1.94}$ \\
    & \nii     & $23.05 \pm 1.96$ \\
    \hline
    \multirow{4}{*}{2}  & \ac{FBP} & $21.99 \pm 1.95$ \\
    & \ac{LS}               & $\mathbf{31.03 \pm 1.88}$ \\
    & \ac{SDF}           & $30.59 \pm 1.91$ \\
    & \nii     & $28.66 \pm 1.93$ \\
    \bottomrule
    \end{tabular}
    \caption{Comparison of the \ac{PSNR} reconstruction metric for \ac{FBP}, \ac{LS}, \ac{SDF} and \nii for two modes of the 2DeteCT dataset. Bold font is used to highlight the best metrics.}
    \label{tab:results_2deteCT_sparse_view}
\end{table}
For Mode1 sinograms, \ac{SDF} provides a 20dB and a 3.9dB improvement against \ac{FBP} and \nii, respectively. It is also better than the \ac{LS}, which ran for 100 iterations but only by a short 0.3dB margin. For Mode2 sinograms, \ac{SDF} provides an 8.6dB and a 1.9dB improvement against \ac{FBP} and \nii, respectively. \ac{LS} remains the best method for high-dose sinograms, with a 0.4dB improvement compared to \ac{SDF}. We show the same slice reconstructed using different methods in App.~\ref{appendix:2deteCT_sparse_view}.

\subsection{\Ac{SDF} as a pre-training step for supervised methods}
We want to measure how good of a pre-training method \ac{SDF} is. To do so, we use the validation subset of the 2DeteCT dataset and fine-tune \ac{SDF} and \nii in a supervised way with the Mode2 golden standard reconstruction. We set the learning rate to $5 \cdot 10^{-6}$. 

For further comparison, we train a neural network similar to the one described in Fig.~\ref{fig:vanilla_CNN} in a supervised fashion from scratch on images reconstructed with \ac{FBP} and \ac{LS}. All hyperparameters remain the same, except the learning rate we set to $10^{-4}$. We present the results in Table~\ref{tab:results_2deteCT_fine_tuning}.

\begin{table}[]
    \centering
    \begin{tabular}{cll}
    \toprule
    Mode & Method & Sparse-View \acs{PSNR} \\
    \midrule
    \multirow{4}{*}{1} & \acs{FBP}     & $27.52 \pm 1.95$ \\
                       & \acs{LS}     & $\mathbf{29.19 \pm 1.90}$ \\
                       & \acs{SDF} & $28.34 \pm 1.95$ \\
                       & \nii    & $27.61 \pm 1.94$ \\
    \hline
    \multirow{4}{*}{2} & \acs{FBP}     & $33.32 \pm 1.93$ \\
                       & \acs{LS}     & $30.59 \pm 1.89$ \\
                       & \acs{SDF} & $\mathbf{34.21 \pm 1.94}$ \\
                       & \nii    & $32.59 \pm 1.92$ \\
    \bottomrule
    \end{tabular}
    \caption{Comparison of the \ac{PSNR} reconstruction metric for \ac{FBP}, \ac{LS}, \ac{SDF} and \nii for two modes of the 2DeteCT dataset. Bold font is used to highlight the best metrics.}
    \label{tab:results_2deteCT_fine_tuning}
\end{table}

The first observation is that the \ac{FBP} benefits the most from supervision. On just 10\% of the training dataset, this simple method yields a 20.8dB improvement for Mode1 and an 11.3dB improvement for Mode2, placing it second best for this mode. Then, the images produced by \ac{LS} are the ones benefiting the less from \ac{CNN} enhancement. Indeed, only 2.6dB is gained for Mode1, and, interestingly, training a \ac{CNN} in a supervised manner on images reconstructed with \ac{LS} diminishes the \ac{PSNR} compared to the original, machine-learning-free reconstruction. For the self-supervised pre-training followed by supervised fine-tuning, we report an improvement of \ac{SDF} by 1.4dB and 3.6dB for Mode1 and Mode2, respectively. As for the \nii method, supervised fine-tuning improves the Mode1 images by 4.6 dB and the Mode2 images by 3.9 dB, closing its gap with \ac{SDF}. Although the comparative improvement yielded by \ac{SDF} over \nii is carried over by supervised fine-tuning, the delta between the two methods closes.

\subsection{\Ac{SDF} applied to sparse sinograms}
We control for the performance of unsupervised methods against angular sparsity, so we train \ac{SDF} and \nii on sinograms divided into 2, 5, 10, 12, 15, 18 and 25 subsets. 

\begin{figure}
    \centering
   \begin{tikzpicture}[
    spy scope={magnification=4, size=2.5cm},
    every spy on node/.style={fill, fill opacity=0.2, text opacity=1},
    every spy in node/.style={draw, ultra thick}
    ]
  \node[inner sep=0pt] (target) at (0,0)
    {
        \includegraphics[width=0.4\textwidth]{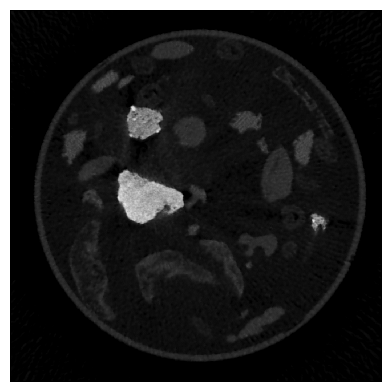}
    };

  \node[inner sep=0pt] (FDK) at (0,-6)
    {
        \includegraphics[width=0.4\textwidth]
            {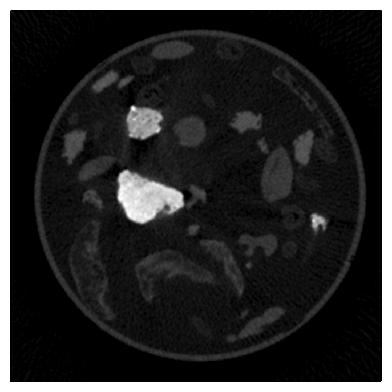}
    };

 \spy [green,magnification=2.5, thick] on (-0.55,0) in node at (4.5,1.5);
  \spy [red,magnification=4] on (-0.65,0.9) in node at (7.5,1.5);
  \spy [orange,magnification=5] on (1.5,-0.3) in node at (4.5,-1.5);
  \spy [blue,magnification=2.5] on (0.45,-0.95) in node at (7.5,-1.5);

  \spy [green,magnification=2.5, thick] on (-0.55,-6) in node at (4.5,-4.5);
  \spy [red,magnification=4] on (-0.65,-5.1) in node at (7.5,-4.5);
  \spy [orange,magnification=5] on (1.5,-6.3) in node at (4.5,-7.5);
  \spy [blue,magnification=2.5] on (0.45,-6.95) in node at (7.5,-7.5);

\end{tikzpicture}
    \caption{Close-up comparison of the same slice reconstructed from 240 measurements by \ac{SDF} (top) and \nii (bottom).}
    \label{fig:subsets_close_up}
\end{figure}

\begin{tikzpicture}
\begin{axis}[
    title={Average \acs{PSNR} on the Test Dataset Against Number of Sinogram Subsets},
    xlabel={Number of Subsets},
    ylabel={PSNR},
    xmin=0, xmax=26,
    ymin=28, ymax=33,
    xtick={2,5,10,12,15,18,25},
    legend pos=north east,
    ymajorgrids=true,
    grid style=dashed,
]

\addplot[
    color=blue,
    mark=+,
    ]
    coordinates {
    (2, 31.53)(5, 30.93)(10, 30.59)(12, 30.62)(15, 31.28)(18, 30.89)(25, 30.21)
    };

\addplot[
    color=orange,
    mark=x,
    ]
    coordinates {
    (2, 28.53)(5, 28.65)(10, 28.66)(12, 28.71)(15, 28.34)(18, 28.20)(25, 28.82)
    };

\legend{\ac{SDF}, \nii}
    
\end{axis}

\end{tikzpicture}
We observe a \ac{PSNR} gap of about 3dB between \ac{SDF} and \nii. These results are surprising. Although we would expect a reduction of the average \ac{PSNR} as the number of subsets increases, i.e. when there are fewer measurements per sinogram, we observe that for 15 subsets, the performance of \ac{SDF} sharply increases by 0.7dB. This means that a sinogram can be sub-sampled from 3600 to 240 angular measurements and still maintain a similar image quality as 1800 measurement sinograms improved with \ac{SDF}. This opens the door for further investigations on the optimal angular subsampling strategy. We show the difference between the same slice reconstructed with \ac{SDF} and \nii in Fig.~\ref{fig:subsets_close_up} and a detailed comparison of the reconstruction's quality for each sparsity level in Fig.~\ref{tab:subsets_full_comparison}.

\begin{table}[]
    \centering
    \begin{tabular}{ccc}
                     & SDF & Noise2Inverse  \\
          2 Subsets  & \includegraphics[scale=0.45]{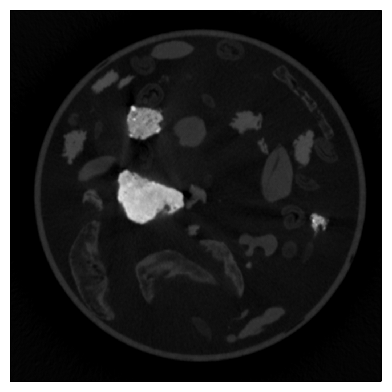} & \includegraphics[scale=0.45]{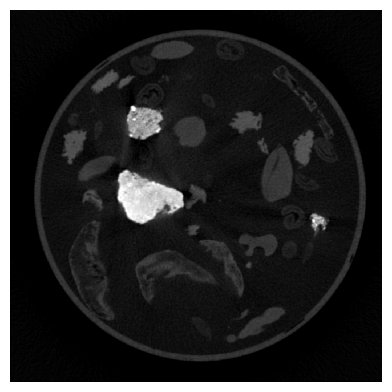} \\
          15 Subsets & \includegraphics[scale=0.45]{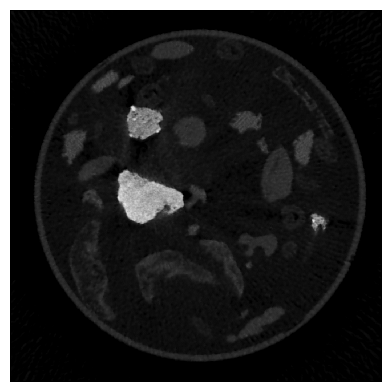} & \includegraphics[scale=0.45]{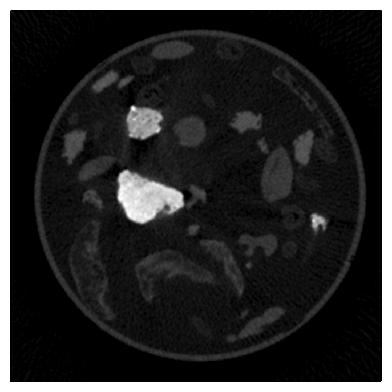} \\
          25 Subsets & \includegraphics[scale=0.45]{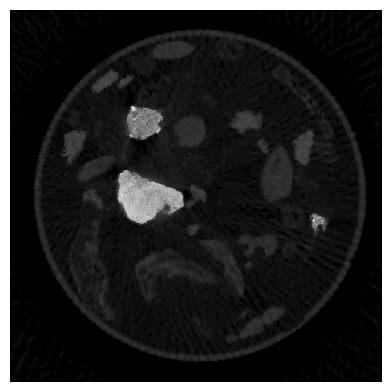} & \includegraphics[scale=0.45]{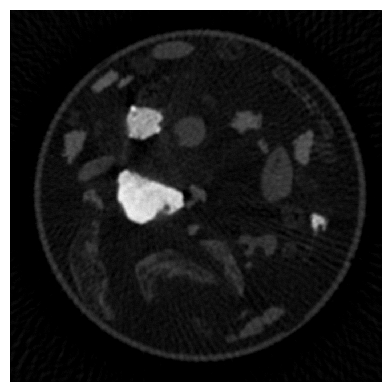} \\

    \end{tabular}
    \caption{Comparison of Images reconstructed with SDF and Noise2Inverse from 2, 15 and 25 subsets. One notable feature of the images reconstructed with SDF is that the high attenuation areas, the lava-stones, remain sharp even when few measurements are available, whereas Noise2Inverse seems to blur the same areas. That SDF feature might be desirable for segmentation purposes.}
    \label{tab:subsets_full_comparison}
\end{table}

\subsection{\Ac{SDF} for Orbital and Angular Subsets of \ac{CBCT} Sinograms}
We experiment on the Walnuts dataset to assess the scalability of our approach to 3D \ac{CBCT}. This dataset comprises 42 \ac{CBCT} scans of walnuts, which we split into 35 samples for training and 7 samples for testing. We have found sample 40 to be artifact-laden, as running an \ac{FDK} on the data yielded a 5dB \ac{PSNR}. We then exclude this specific sample from the test set, bringing it to 6 walnuts. We divide the sinogram into 36 subsets, 3 orbitals and 12 angular, and use a batch size of one. As such, the input of our network is the \ac{FDK} of a sample reconstructed from 100 measurements evenly distributed across $2\pi$ radians. Each measurement is a $972 \times 768$ matrix. Our only comparison is the \ac{FDK} of each sinogram subset. We evaluate the impact of each orbital subset on the reconstruction's quality by presenting a per-orbit result in Table~\ref{tab:results_walnuts}.

\begin{table}[]
    \centering
    \begin{tabular}{llll}
    \toprule
    Method Name & Orbit 1 \acs{PSNR} & Orbit 2 \acs{PSNR} & Orbit 3 \acs{PSNR} \\
    \midrule    
    \acs{FDK} & $28.53 \pm 2.31$ & $28.72 \pm 2.34$ & $28.70 \pm 2.34$ \\
    \acs{SDF} & $36.21 \pm 2.35$ & $37.05 \pm 2.33$ & $37.03 \pm 2.39$ \\
    \bottomrule
    \end{tabular}
    \caption{Results of experiments on angular and orbital subsets of the Walnuts dataset.}
\label{tab:results_walnuts}
\end{table}

\begin{figure}
    \centering
   \begin{tikzpicture}[
    spy scope={magnification=4, size=2.5cm},
    every spy on node/.style={fill, fill opacity=0.2, text opacity=1},
    every spy in node/.style={draw, ultra thick}
    ]
  \node[inner sep=0pt] (target) at (0,0)
    {\includegraphics[width=0.4\textwidth]{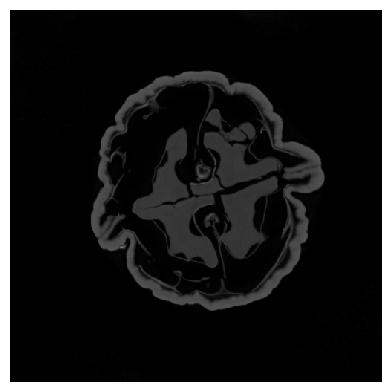}};

  \node[inner sep=0pt] (FDK) at (0,-6)
    {\includegraphics[width=0.4\textwidth]{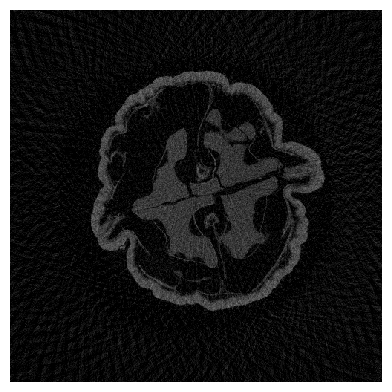}};

  \node[inner sep=0pt] (method) at (0,-12)
    {\includegraphics[width=0.4\textwidth]{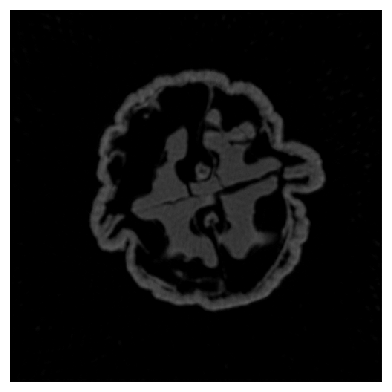}};  

  \spy [red,magnification=4] on (0.1,0.4) in node at (4.5,0);
  \spy [orange,magnification=5] on (0.2, -0.35) in node at (7.5,0);

   \spy [red,magnification=4] on (0.1,-5.6) in node at (4.5,-6);
  \spy [orange,magnification=5] on (0.2, -6.35) in node at (7.5,-6);

   \spy [red,magnification=4] on (0.1,-11.6) in node at (4.5,-12);
  \spy [orange,magnification=5] on (0.2, -12.35) in node at (7.5,-12);
\end{tikzpicture}
    \caption{Central Slice of Walnut 39 reconstructed with \ac{LS} on full sinogram, \ac{FDK} and +\ac{SDF} on subsampled sinogram.}
    \label{fig:walnuts_qualitative}
\end{figure}

We report a 8.1dB \ac{PSNR} improvement compared to the base, \ac{FDK}. We also observe that images reconstructed from the first orbit using both methods are consistently worse than images reconstructed from orbits 2 and 3. Although these results are encouraging, we want to underline that \ac{SDF} seems to smooth out fine details of the walnuts, as underlined in Fig.~\ref{fig:walnuts_qualitative}.

\section{Conclusion}
We present \ac{SDF}, a novel framework for training image denoisers in a self-supervised fashion. We establish its advantage over the other self-supervised reconstruction method, \nii,  for low and high photon counts. This might come from the fact that \nii was not designed to address sparse-view \ac{CT} artifacts and that the $1:X$ training strategy they delineated is, according to them, inferior to its counterpart $X:1$. We showed that \ac{SDF} was a suitable pre-training strategy, as supervised fine-tuning of the pre-trained network on a fraction of the dataset maintains the advantage on \nii. However, the gap closes for the studied angular sparsity level. We then study how robust against angular sparsity \ac{SDF} is compared to \nii. We show that higher angular sparsity appears to be beneficial to \ac{SDF}, a regime where \nii struggles. Then, we demonstrate the potential of \ac{SDF} on \ac{CBCT} by combining orbital and angular sparsity by downsampling the input sinogram by a factor of 36 whilst maintaining a 37dB \ac{PSNR} similarity to the full-view image reconstructed with \ac{LS}. This establishes solid ground for the adoption of \ac{SDF} as a pre-training strategy, as the network can be trained on full-view data acquired on as little as 35 different samples and then used in settings with drastically increased angular sparsity, speeding up the acquisition process substantially. 

\clearpage

\paragraph{Acknowledgements}
We thank Nathan Kutz for his talk at KTH focus period on SciML, which initiated this investigation.
OÖ and EV acknowledge support from Swedish Energy Agency grant P2022-00286 and FORMAS grant 2022-00469.
AH acknowledges support from DigitalFutures Scholar-in-Residence grant KTH-RPROJ-0146472
2.

\printbibliography

@misc{hamamci_developing_2024,
	title = {Developing {Generalist} {Foundation} {Models} from a {Multimodal} {Dataset} for {3D} {Computed} {Tomography}},
	url = {https://arxiv.org/abs/2403.17834},
	author = {Hamamci, Ibrahim Ethem and Er, Sezgin and Almas, Furkan and Simsek, Ayse Gulnihan and Esirgun, Sevval Nil and Dogan, Irem and Dasdelen, Muhammed Furkan and Durugol, Omer Faruk and Wittmann, Bastian and Amiranashvili, Tamaz and Simsar, Enis and Simsar, Mehmet and Erdemir, Emine Bensu and Alanbay, Abdullah and Sekuboyina, Anjany and Lafci, Berkan and Bluethgen, Christian and Ozdemir, Mehmet Kemal and Menze, Bjoern},
	year = {2024},
	note = {\_eprint: 2403.17834},
}

@misc{devlin_bert_2019,
	title = {{BERT}: {Pre}-training of {Deep} {Bidirectional} {Transformers} for {Language} {Understanding}},
	url = {https://arxiv.org/abs/1810.04805},
	author = {Devlin, Jacob and Chang, Ming-Wei and Lee, Kenton and Toutanova, Kristina},
	year = {2019},
	note = {\_eprint: 1810.04805},
}

@article{aarle_fast_2016,
	title = {Fast and flexible {X}-ray tomography using the {ASTRA} toolbox},
	volume = {24},
	url = {https://opg.optica.org/oe/abstract.cfm?URI=oe-24-22-25129},
	doi = {10.1364/OE.24.025129},
	number = {22},
	journal = {Opt. Express},
	author = {Aarle, Wim van and Palenstijn, Willem Jan and Cant, Jeroen and Janssens, Eline and Bleichrodt, Folkert and Dabravolski, Andrei and Beenhouwer, Jan De and Batenburg, K. Joost and Sijbers, Jan},
	month = oct,
	year = {2016},
	note = {Publisher: Optica Publishing Group},
	pages = {25129--25147},
}

@article{hendriksen_tomosipo_2021,
	title = {Tomosipo: fast, flexible, and convenient {3D} tomography for complex scanning geometries in {Python}},
	volume = {29},
	url = {https://opg.optica.org/oe/abstract.cfm?URI=oe-29-24-40494},
	doi = {10.1364/OE.439909},
	number = {24},
	journal = {Opt. Express},
	author = {Hendriksen, Allard A. and Schut, Dirk and Palenstijn, Willem Jan and Viganó, Nicola and Kim, Jisoo and Pelt, Daniël M. and Leeuwen, Tristan van and Batenburg, K. Joost},
	month = nov,
	year = {2021},
	note = {Publisher: Optica Publishing Group},
	pages = {40494--40513},
}

@misc{biguri_learned_nodate,
	title = {Learned {Iterative} {Optimization} {Networks} ({LION})},
	url = {https://github.com/CambridgeCIA/LION/tree/main},
	author = {Biguri, Ander},
}

@misc{paszke_pytorch_2019,
	title = {{PyTorch}: {An} {Imperative} {Style}, {High}-{Performance} {Deep} {Learning} {Library}},
	url = {https://arxiv.org/abs/1912.01703},
	author = {Paszke, Adam and Gross, Sam and Massa, Francisco and Lerer, Adam and Bradbury, James and Chanan, Gregory and Killeen, Trevor and Lin, Zeming and Gimelshein, Natalia and Antiga, Luca and Desmaison, Alban and Köpf, Andreas and Yang, Edward and DeVito, Zach and Raison, Martin and Tejani, Alykhan and Chilamkurthy, Sasank and Steiner, Benoit and Fang, Lu and Bai, Junjie and Chintala, Soumith},
	year = {2019},
	note = {\_eprint: 1912.01703},
}

@misc{chen_robust_2022,
	title = {Robust {Equivariant} {Imaging}: a fully unsupervised framework for learning to image from noisy and partial measurements},
	url = {https://arxiv.org/abs/2111.12855},
	author = {Chen, Dongdong and Tachella, Julián and Davies, Mike E.},
	year = {2022},
	note = {\_eprint: 2111.12855},
}

@article{unal_proj2proj_2024,
	title = {{Proj2Proj}: self-supervised low-dose {CT} reconstruction.},
	volume = {10},
	copyright = {© 2024 Unal et al.},
	issn = {2376-5992},
	doi = {10.7717/peerj-cs.1849},
	language = {eng},
	journal = {PeerJ. Computer science},
	author = {Unal, Mehmet Ozan and Ertas, Metin and Yildirim, Isa},
	year = {2024},
	pmid = {38435612},
	pmcid = {PMC10909204},
	note = {Place: United States},
	pages = {e1849},
}

@misc{kosomaa_simulator-based_2023,
	title = {Simulator-{Based} {Self}-{Supervision} for {Learned} {3D} {Tomography} {Reconstruction}},
	url = {https://arxiv.org/abs/2212.07431},
	author = {Kosomaa, Onni and Laine, Samuli and Karras, Tero and Aittala, Miika and Lehtinen, Jaakko},
	year = {2023},
	note = {\_eprint: 2212.07431},
}

@article{adler_task_2022,
	title = {Task adapted reconstruction for inverse problems},
	volume = {38},
	number = {7},
	journal = {Inverse Problems},
	author = {Adler, Jonas and Lunz, Sebastian and Verdier, Olivier and Schönlieb, Carola-Bibiane and Öktem, Ozan},
	year = {2022},
	note = {Publisher: IOP Publishing},
	pages = {075006},
}

@article{eijnatten_3d_2021,
	title = {{3D} deformable registration of longitudinal abdominopelvic {CT} images using unsupervised deep learning},
	volume = {208},
	issn = {0169-2607},
	url = {https://www.sciencedirect.com/science/article/pii/S0169260721003357},
	doi = {https://doi.org/10.1016/j.cmpb.2021.106261},
	journal = {Computer Methods and Programs in Biomedicine},
	author = {Eijnatten, Maureen van and Rundo, Leonardo and Batenburg, K. Joost and Lucka, Felix and Beddowes, Emma and Caldas, Carlos and Gallagher, Ferdia A. and Sala, Evis and Schönlieb, Carola-Bibiane and Woitek, Ramona},
	year = {2021},
	pages = {106261},
}

@article{valat_empirical_2024,
	title = {Empirical evidence of the task-adapted reconstruction framework for joint {CT} reconstruction and segmentation},
	volume = {2},
	number = {3},
	journal = {Applied Mathematics for Modern Challenges},
	author = {Valat, Emilien and Biguri, Ander and Sanchez, Lorena Escudero and McCague, Cathal and Öktem, Ozan and Schönlieb, Carola-Bibiane},
	year = {2024},
	note = {Publisher: Applied Mathematics for Modern Challenges},
	pages = {287--300},
}

@inproceedings{genzel_near-exact_2022,
	title = {Near-exact recovery for tomographic inverse problems via deep learning},
	booktitle = {International {Conference} on {Machine} {Learning}},
	publisher = {PMLR},
	author = {Genzel, Martin and Gühring, Ingo and Macdonald, Jan and März, Maximilian},
	year = {2022},
	pages = {7368--7381},
}

@article{leuschner_quantitative_2021,
	title = {Quantitative {Comparison} of {Deep} {Learning}-{Based} {Image} {Reconstruction} {Methods} for {Low}-{Dose} and {Sparse}-{Angle} {CT} {Applications}},
	volume = {7},
	issn = {2313-433X},
	url = {https://www.mdpi.com/2313-433X/7/3/44},
	doi = {10.3390/jimaging7030044},
	number = {3},
	journal = {Journal of Imaging},
	author = {Leuschner, Johannes and Schmidt, Maximilian and Ganguly, Poulami Somanya and Andriiashen, Vladyslav and Coban, Sophia Bethany and Denker, Alexander and Bauer, Dominik and Hadjifaradji, Amir and Batenburg, Kees Joost and Maass, Peter and van Eijnatten, Maureen},
	year = {2021},
}

@article{arridge_solving_2019,
	title = {Solving inverse problems using data-driven models},
	volume = {28},
	doi = {10.1017/S0962492919000059},
	journal = {Acta Numerica},
	author = {Arridge, Simon and Maass, Peter and Öktem, Ozan and Schönlieb, Carola-Bibiane},
	year = {2019},
	pages = {1--174},
}

@article{kiss_2detect_2023,
	title = {{2DeteCT} -- {A} large {2D} expandable, trainable, experimental {Computed} {Tomography} dataset for machine learning},
	volume = {10},
	issn = {2052-4463},
	url = {http://arxiv.org/abs/2306.05907},
	doi = {10.1038/s41597-023-02484-6},
	number = {1},
	urldate = {2024-06-27},
	journal = {Scientific Data},
	author = {Kiss, Maximilian B. and Coban, Sophia B. and Batenburg, K. Joost and van Leeuwen, Tristan and Lucka, Felix},
	month = sep,
	year = {2023},
	note = {arXiv:2306.05907 [cs, eess]},
	pages = {576},
}

@article{der_sarkissian_cone-beam_2019,
	title = {A cone-beam {X}-ray computed tomography data collection designed for machine learning},
	volume = {6},
	copyright = {2019 The Author(s)},
	issn = {2052-4463},
	url = {https://www.nature.com/articles/s41597-019-0235-y},
	doi = {10.1038/s41597-019-0235-y},
	language = {en},
	number = {1},
	urldate = {2024-07-10},
	journal = {Scientific Data},
	author = {Der Sarkissian, Henri and Lucka, Felix and van Eijnatten, Maureen and Colacicco, Giulia and Coban, Sophia Bethany and Batenburg, Kees Joost},
	month = oct,
	year = {2019},
	pages = {215},
}

@article{hendriksen_noise2inverse_2020,
	title = {{Noise2Inverse}: {Self}-{Supervised} {Deep} {Convolutional} {Denoising} for {Tomography}.},
	volume = {6},
	url = {http://dblp.uni-trier.de/db/journals/tci/tci6.html#HendriksenPB20},
	journal = {IEEE Trans. Computational Imaging},
	author = {Hendriksen, Allard A. and Pelt, Daniël Maria and Batenburg, Kees Joost},
	year = {2020},
	pages = {1320--1335},
}

\pagebreak

\begin{appendices}
\section{2DeteCT - Sparse-View Images}
\label{appendix:2deteCT_sparse_view}
Throughout this appendix, we always refer to the same slice for Mode1 and Mode2, and always highlight the same areas for a comprehensive comparison.
\begin{figure}[h]
    \centering
    \begin{tikzpicture}[
    spy scope={magnification=4, size=2.5cm},
    every spy on node/.style={fill, fill opacity=0.2, text opacity=1},
    every spy in node/.style={draw, ultra thick}
    ]
  \node[inner sep=0pt] (image) at (0,0)
    {\includegraphics[width=0.45\textwidth]{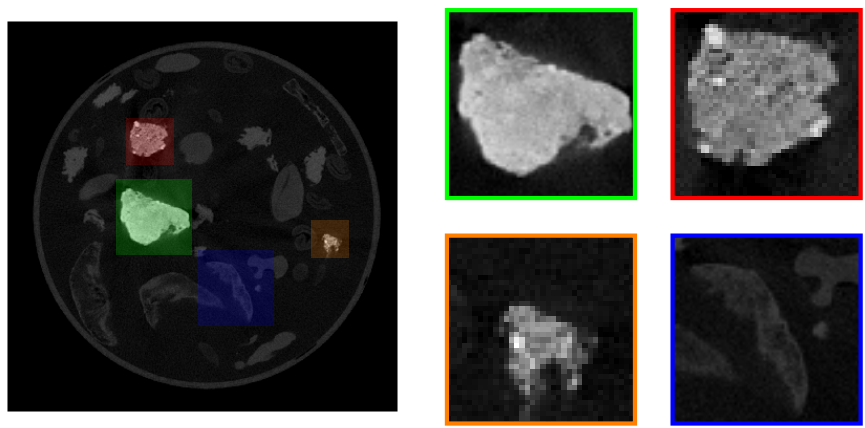}};
  \spy [green,magnification=2.5] on (-0.65,0) in node at (4.5,1.5);
  \spy [red,magnification=4] on (-0.7,1) in node at (7.5,1.5);
  \spy [orange,magnification=5] on (1.7,-0.3) in node at (4.5,-1.5);
  \spy [blue,magnification=2.5] on (0.45,-0.95) in node at (7.5,-1.5);
\end{tikzpicture}
    \caption{Slice of the 2DeteCT Mode2 Test dataset.}
    \label{fig:mode2_target}
\end{figure}

\begin{figure}[h]
    \centering
   \begin{tikzpicture}[
    spy scope={magnification=4, size=2.5cm},
    every spy on node/.style={fill, fill opacity=0.2, text opacity=1},
    every spy in node/.style={draw, ultra thick}
    ]
  \node[inner sep=0pt] (image) at (0,0)
    {\includegraphics[width=0.45\textwidth]{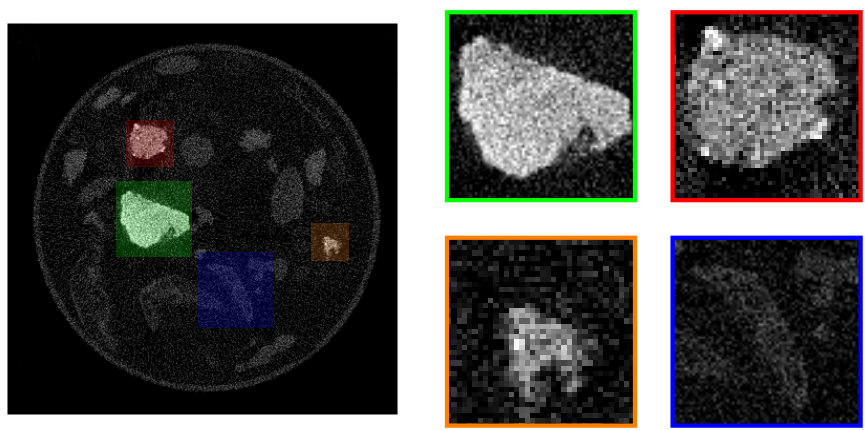}};

  \spy [green,magnification=2.5, thick] on (-0.65,0) in node at (4.5,1.5);
  \spy [red,magnification=4] on (-0.7,1) in node at (7.5,1.5);
  \spy [orange,magnification=5] on (1.7,-0.3) in node at (4.5,-1.5);
  \spy [blue,magnification=2.5] on (0.45,-0.95) in node at (7.5,-1.5);
\end{tikzpicture}
    \caption{Slice of the 2DeteCT Mode1 Test dataset. }
    \label{fig:mode1_target}
\end{figure}

\begin{figure}
    \centering
   \begin{tikzpicture}[
    spy scope={magnification=4, size=2.5cm},
    every spy on node/.style={fill, fill opacity=0.2, text opacity=1},
    every spy in node/.style={draw, ultra thick}
    ]
  \node[inner sep=0pt] (image) at (0,0)
    {\includegraphics[width=0.45\textwidth]{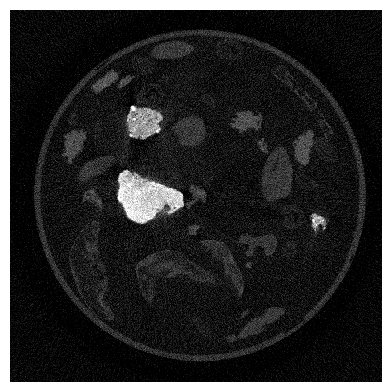}};

  \spy [green,magnification=2.5, thick] on (-0.65,0) in node at (4.5,1.5);
  \spy [red,magnification=4] on (-0.7,1) in node at (7.5,1.5);
  \spy [orange,magnification=5] on (1.7,-0.3) in node at (4.5,-1.5);
  \spy [blue,magnification=2.5] on (0.45,-0.95) in node at (7.5,-1.5);
\end{tikzpicture}
    \caption{Slice of the 2DeteCT Mode2 Test dataset reconstructed with \ac{FBP}.}
    \label{fig:mode2_FBP}
\end{figure}

\begin{figure}
    \centering
   \begin{tikzpicture}[
    spy scope={magnification=4, size=2.5cm},
    every spy on node/.style={fill, fill opacity=0.2, text opacity=1},
    every spy in node/.style={draw, ultra thick}
    ]
  \node[inner sep=0pt] (image) at (0,0)
    {\includegraphics[width=0.45\textwidth]{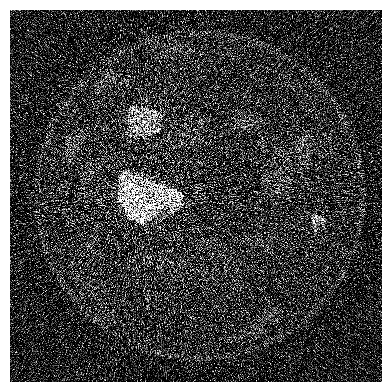}};

  \spy [green,magnification=2.5, thick] on (-0.65,0) in node at (4.5,1.5);
  \spy [red,magnification=4] on (-0.7,1) in node at (7.5,1.5);
  \spy [orange,magnification=5] on (1.7,-0.3) in node at (4.5,-1.5);
  \spy [blue,magnification=2.5] on (0.45,-0.95) in node at (7.5,-1.5);
\end{tikzpicture}
    \caption{Slice of the 2DeteCT Mode1 Test dataset reconstructed with \ac{FBP}. }
    \label{fig:mode1_FBP}
\end{figure}

\begin{figure}
    \centering
   \begin{tikzpicture}[
    spy scope={magnification=4, size=2.5cm},
    every spy on node/.style={fill, fill opacity=0.2, text opacity=1},
    every spy in node/.style={draw, ultra thick}
    ]
  \node[inner sep=0pt] (image) at (0,0)
    {\includegraphics[width=0.45\textwidth]{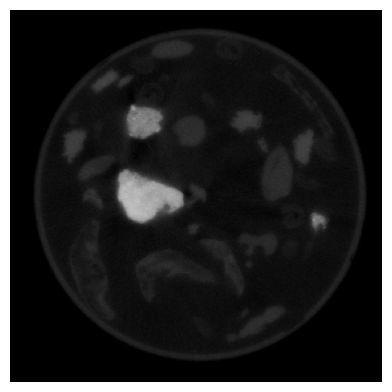}};

  \spy [green,magnification=2.5, thick] on (-0.65,0) in node at (4.5,1.5);
  \spy [red,magnification=4] on (-0.7,1) in node at (7.5,1.5);
  \spy [orange,magnification=5] on (1.7,-0.3) in node at (4.5,-1.5);
  \spy [blue,magnification=2.5] on (0.45,-0.95) in node at (7.5,-1.5);
\end{tikzpicture}
    \caption{Slice of the 2DeteCT Mode2 Test dataset reconstructed with \ac{LS}.}
    \label{fig:mode2_AGD}
\end{figure}

\begin{figure}
    \centering
   \begin{tikzpicture}[
    spy scope={magnification=4, size=2.5cm},
    every spy on node/.style={fill, fill opacity=0.2, text opacity=1},
    every spy in node/.style={draw, ultra thick}
    ]
  \node[inner sep=0pt] (image) at (0,0)
    {\includegraphics[width=0.45\textwidth]{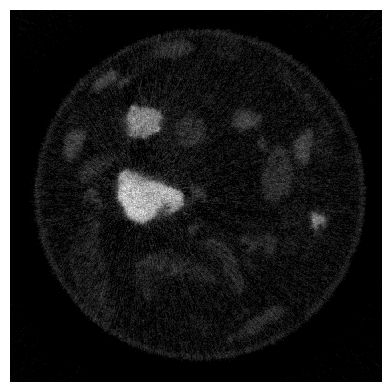}};

  \spy [green,magnification=2.5, thick] on (-0.65,0) in node at (4.5,1.5);
  \spy [red,magnification=4] on (-0.7,1) in node at (7.5,1.5);
  \spy [orange,magnification=5] on (1.7,-0.3) in node at (4.5,-1.5);
  \spy [blue,magnification=2.5] on (0.45,-0.95) in node at (7.5,-1.5);
\end{tikzpicture}
    \caption{Slice of the 2DeteCT Mode1 Test dataset reconstructed with \ac{LS}.}
    \label{fig:mode1_AGD}
\end{figure}

\begin{figure}
    \centering
   \begin{tikzpicture}[
    spy scope={magnification=4, size=2.5cm},
    every spy on node/.style={fill, fill opacity=0.2, text opacity=1},
    every spy in node/.style={draw, ultra thick}
    ]
  \node[inner sep=0pt] (image) at (0,0)
    {\includegraphics[width=0.45\textwidth]{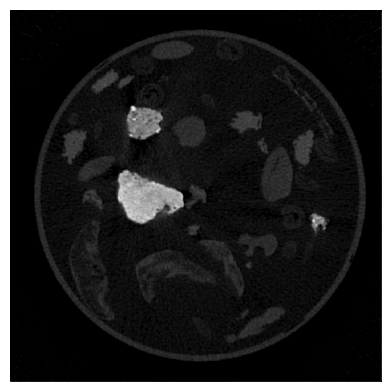}};

  \spy [green,magnification=2.5, thick] on (-0.65,0) in node at (4.5,1.5);
  \spy [red,magnification=4] on (-0.7,1) in node at (7.5,1.5);
  \spy [orange,magnification=5] on (1.7,-0.3) in node at (4.5,-1.5);
  \spy [blue,magnification=2.5] on (0.45,-0.95) in node at (7.5,-1.5);
\end{tikzpicture}
    \caption{Slice of the 2DeteCT Mode2 Test dataset reconstructed with \ac{SDF}.}
    \label{fig:mode2_method}
\end{figure}

\begin{figure}
    \centering
   \begin{tikzpicture}[
    spy scope={magnification=4, size=2.5cm},
    every spy on node/.style={fill, fill opacity=0.2, text opacity=1},
    every spy in node/.style={draw, ultra thick}
    ]
  \node[inner sep=0pt] (image) at (0,0)
    {\includegraphics[width=0.45\textwidth]{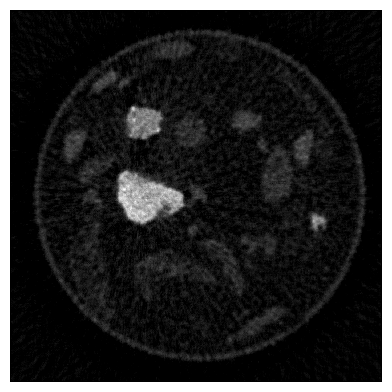}};

  \spy [green,magnification=2.5, thick] on (-0.65,0) in node at (4.5,1.5);
  \spy [red,magnification=4] on (-0.7,1) in node at (7.5,1.5);
  \spy [orange,magnification=5] on (1.7,-0.3) in node at (4.5,-1.5);
  \spy [blue,magnification=2.5] on (0.45,-0.95) in node at (7.5,-1.5);
\end{tikzpicture}
    \caption{Slice of the 2DeteCT Mode1 Test dataset reconstructed with \ac{SDF}.}
    \label{fig:mode1_method}
\end{figure}

\begin{figure}
    \centering
   \begin{tikzpicture}[
    spy scope={magnification=4, size=2.5cm},
    every spy on node/.style={fill, fill opacity=0.2, text opacity=1},
    every spy in node/.style={draw, ultra thick}
    ]
  \node[inner sep=0pt] (image) at (0,0)
    {\includegraphics[width=0.45\textwidth]{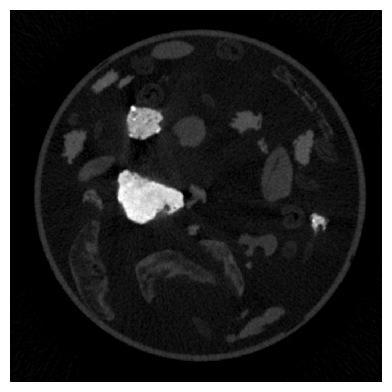}};

  \spy [green,magnification=2.5, thick] on (-0.65,0) in node at (4.5,1.5);
  \spy [red,magnification=4] on (-0.7,1) in node at (7.5,1.5);
  \spy [orange,magnification=5] on (1.7,-0.3) in node at (4.5,-1.5);
  \spy [blue,magnification=2.5] on (0.45,-0.95) in node at (7.5,-1.5);
\end{tikzpicture}
    \caption{Slice of the 2DeteCT Mode2 Test dataset reconstructed with \nii. }
    \label{fig:mode2_nii}
\end{figure}

\begin{figure}
    \centering
   \begin{tikzpicture}[
    spy scope={magnification=4, size=2.5cm},
    every spy on node/.style={fill, fill opacity=0.2, text opacity=1},
    every spy in node/.style={draw, ultra thick}
    ]
  \node[inner sep=0pt] (image) at (0,0)
    {\includegraphics[width=0.45\textwidth]{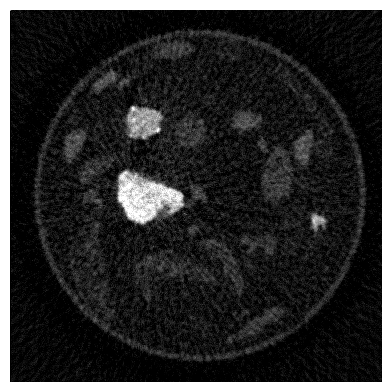}};

  \spy [green,magnification=2.5, thick] on (-0.65,0) in node at (4.5,1.5);
  \spy [red,magnification=4] on (-0.7,1) in node at (7.5,1.5);
  \spy [orange,magnification=5] on (1.7,-0.3) in node at (4.5,-1.5);
  \spy [blue,magnification=2.5] on (0.45,-0.95) in node at (7.5,-1.5);
\end{tikzpicture}
    \caption{Slice of the 2DeteCT Mode1 Test dataset reconstructed with \nii.}
    \label{fig:mode1_nii}
\end{figure}

\end{appendices}

\end{document}